\documentclass[runningheads]{llncs}
\usepackage{alltt}
\usepackage{xcolor}
\usepackage[T1]{fontenc}
\usepackage{hyperref}
\usepackage{amsmath}
\usepackage{bm}
\usepackage{float}
\usepackage{graphicx}
\usepackage{subcaption}

\begin{document}
\title{DataSafe: Copyright Protection with PUF Watermarking and Blockchain Tracking}
\author{
Xiaolong Xue\inst{1}%\orcidID{0000-1111-2222-3333} 
\and
Guangyong Shang\inst{2} 
\and
Zhen Ma\inst{2} 
\and
Minghui Xu\inst{1*}%\orcidID{0000-0003-3675-3461} 
\and
Hechuan Guo\inst{1}%\orcidID{0000-0003-0397-1382} 
\and
Kun Li\inst{1}%\orcidID{0000-0002-8305-0841} 
\and
Xiuzhen Cheng\inst{1}%\orcidID{0000-0001-5912-4647}
}
\authorrunning{
X. Xue et al.
}
\institute{
Shandong University, Jinan, China\\
\email{\{xiaolongxue, ghc\}@mail.sdu.edu.cn, \{mhxu, kunli, xzcheng\}@sdu.edu.cn}\\
\and
Inspur Yunzhou Industrial Internet Co., Ltd., Jinan, China\\
\email{\{shangguangyong, mazhenrj\}@inspur.com}\\
* Corresponding author: Minghui Xu
Springer Heidelberg, Tiergartenstr. 17, 69121 Heidelberg, Germany
\email{lncs@springer.com}\\
}
\maketitle              % typeset the header of the contribution
\begin{abstract}
Digital watermarking methods are commonly used to safeguard digital media copyrights by confirming ownership and deterring unauthorized use. However, without reliable third-party oversight, these methods risk security vulnerabilities during watermark extraction. Furthermore, digital media lacks tangible ownership attributes, posing challenges for secure copyright transfer and tracing. This study introduces DataSafe, a copyright protection scheme that combines physical unclonable functions (PUFs) and blockchain technology. PUF devices use their unique fingerprints for blockchain registration. Subsequently, these devices incorporate invisible watermarking techniques to embed digital watermarks into media for copyright protection. The watermark verification process is confined within the devices, preserving confidentiality during extraction, validating identities during copyright exchanges, and facilitating blockchain-based traceability of copyright transfers. The implementation of a prototype system on the LPC55S69-EVK development board is detailed, illustrating the practicality and effectiveness of the proposed solution.
\keywords{Data security \and Copyright protection \and Watermarks \and Physical unclonable functions \and Blockchain.}
\end{abstract}
\section{Introduction}
\label{sec:intro}
In recent years, the adoption of digital technologies has become a crucial factor driving economic growth and social advancement. Copyright owners and creators share various literary and artistic works as digital media across users' electronic devices. While the lossless distribution of digital media promotes knowledge sharing, it also introduces copyright challenges, including difficulties in verifying ownership and holding infringers accountable. A widely used solution for these issues is applying  watermarks in digital media. The design of digital watermarks aims to achieve robustness, invisibility, and security \cite{lin2011digital} \cite{zhu2008novel} \cite{xiang2008invariant}. However, this approach often overlooks conflicts related to copyright verification: (1) To ensure reliable copyright verification of digital watermarks, the extraction key of the watermark must be public, which inherently risks exposing secrets. (2) Traditional digital watermarks, often based on pseudo-random sequences or binary images, which are easily imitated and used for infringement behavior. (3) Copyright disputes are frequent during transfers, and solely relying on digital watermarks can result in unresolved conflicts, particularly in peer-to-peer distribution networks.

To tackle these challenges, various solutions have been proposed. Dutta et al. \cite{dutta2013efficient} suggested using iris features as digital watermarks embedded in media files. By integrating biometric features as labels in digital media, the issue of watermark counterfeiting can be mitigated. Furthermore, a blockchain service architecture for recording, protecting, verifying, and tracing the registration and transactions of original works has been proposed in \cite{zhu2021using}. Jiang et al. \cite{jiang2020research} examined the advantages of blockchain technology for copyright protection, proposing a lightweight and cost-effective copyright protection system. These methods utilize blockchain evidence to record copyright transfers. Nonetheless, challenges remain because when transferring copyright with others, one can forge their identity or use their identity for framing.

To establish a secure connection between digital media and copyright holders, it is essential to accurately identify the copyright holder while preventing fraudulent activities. Although biometric features provides a potential solution, it still requires the involvement of trusted third parties. Another promising approach for establishing physical binding is the use of physical unclonable functions (PUFs), which are hardware-based security primitives. These rely on the unique physical properties inherent in each circuit, resulting from uncontrollable variations during the manufacturing process \cite{feiten2015improving}. Because these properties cannot be easily replicated, they can be used to generate unique device fingerprints. Building on the use of PUF-based device fingerprints, this paper presents DataSafe, a copyright protection scheme that manages copyright transfer and traceability with PUF and blockchain. This system leverages blockchain technology to enhance security and transparency. The contributions of this paper are as follows:

\begin{enumerate}
    \item By integrating physical unclonable functions (PUFs) as device fingerprints into digital watermarking, we establish physical ownership of digital media copyrights.
    \item We propose a secure digital watermark extraction process utilizing PUF devices to safeguard against secret leakage during copyright verification.
    \item To ensure authenticity during copyright transfers and prevent identity forgery, the proposed system leverages blockchain technology to register PUF devices, facilitating secure information exchange between trading parties.
    \item This approach integrates secure digital watermark extraction with blockchain storage of copyright transfer information, enabling traceability throughout the ownership history.
\end{enumerate}

The structure of this paper is outlined as follows: In Section \ref{sec:related}, we examined related works. In Section \ref{sec:design}, we present our DataSafe design. In Section \ref{sec:prototype}, we establish a prototype founded on the proposed architecture and demonstrate experimental outcomes. Finally, in Section \ref{sec:conclusion}, we summarize this paper.

\section{Related Work}
\label{sec:related}
In this section, we examine pertinent studies concerning device fingerprinting employing PUF, copyright safeguarding through device watermarking, and copyright safeguarding via blockchain, along with associated enhancements in these domains.

\subsection{Device Fingerprinting with PUF}
Embedded devices pose a unique challenge for secure key storage. Unlike traditional systems, attackers can physically access the circuits, potentially compromising keys stored in non-volatile memory (as shown in~\cite{shamsoshoara2020survey}).
Kerckhoff's principle dictates that the security of a cryptosystem should rely solely on the secrecy of the key. Modern systems adhere to this principle by making everything public except the key. To address this vulnerability in embedded devices, Tuyls et al.~\cite{tuyls2007secure} proposed extracting keys from physical properties of the integrated circuits themselves. This eliminates the need for storing keys in memory, mitigating memory-based attacks and guaranteeing device uniqueness and authenticity. This approach can be extended to strengthen key storage in copyright management systems.
PUFs can serve as unique device fingerprints. Existing research explores integrating PUFs into hardware accelerators to protect neural network models~\cite{dorfmeister2024puf,guo2018puf}. However, these methods are limited to specific devices and are not directly applicable to safeguarding digital media copyrights.

\subsection{Copyright Protection with Digital Watermarking}
Watermarks are subtly embedded within the non-essential bits of digital content, altering the data without compromising the user's experience  \cite{fridrich2002lossless}. This process essentially weaves covert data into the original material.
Several studies have explored the use of biometric features as watermarks. Dutta et al. proposed incorporating iris traits into media files \cite{dutta2009blind,dutta2013efficient}, while Wojtowicz et al. \cite{wojtowicz2016digital} advocated for integrating both fingerprint and iris features in digital images . While these methods aim to address forgeable watermarks with biometrics, using biometric features for copyright protection raises concerns. Asserting copyright through biometric features could unintentionally reveal sensitive personal information, and it doesn't fully prevent copyright violations.

% \old{To mitigate the risk of inadvertent disclosure of secrets, we restrict the embedding and extraction of watermarks to devices that are embedded with device fingerprints, ensuring that only those devices can perform watermark-related operations. We utilize a spatial domain embedding approach to directly incorporate the watermark into the excess bits of digital media. While this technique may lack robustness, it provides a significant capacity for embedding and is computationally efficient, making it suitable for devices with limited power. To improve robustness, we apply signature processing to the watermark data during the embedding phase and utilize a random number generator to determine the placement of the watermark data.}

\subsection{Data Protection with Blockchain}

Blockchain technology offers a novel decentralized architecture for distributed applications~\cite{xu2024exploring,cheng2023adaptive}. This technology flourishes alongside smart contracts, self-executing agreements that automate the fulfillment of conditions and streamline processes for all involved parties. Furthermore, blockchain serves as an immutable ledger, akin to a tamper-proof bulletin board, ensuring robust traceability of information~\cite{xu2022curb,xu2022spdl,filedes}.
Lu et al.~\cite{lu2019blockchain} have proposed a scheme leveraging blockchain for the management of digital rights for design works. However, this method necessitates users to submit their private keys to the application, which could be a vulnerability as the decryption program might potentially misappropriate the private key. On a similar note, Ma et al.~\cite{ma2018new} suggested a scheme combining blockchain and watermarking to detect the misuse of images online. Yet, they did not sufficiently address the issue of identity fraud in copyright transfers, which limits the ability to accurately pinpoint copyright violators during the tracing process. Liu et al.~\cite{tbac} have used blockchain to supervise and control tangible devices and data, thus enhancing the dependability of management systems. TEMS~\cite{tems} adopts TEEs to protect the data on-chaining process. 
However, these works do not provide a mechanism to prevent unauthorized data redistribution once the data has been leaked from the system, lacking a watermarking solution to address this issue.

This paper explores a system designed to log copyright transfers on a blockchain and to produce watermarks with the aid of trusted device signatures. It captures the process by which digital media copyrights are transferred from sellers to buyers, establishing comprehensive and auditable records for these transfer transactions via smart contracts. By incorporating the physical ownership of digital media copyrights as detailed in this paper, it is possible to achieve a full traceability of copyright on the blockchain.

% \subsection{Security Assumptions}
% In this paper, we make the following assumptions.

% \begin{enumerate}
%     \item The embedded devices involved in copyright transfer have enabled PUF and PUF is bound to specific devices.
%     \item Any attempt to tamper with the PUF will alter the device's behavior and render the PUF useless.
%     \item There is no access to communication between the device's microcontroller and its PUF, meaning we do not consider the issue of side channel attacks.
% \end{enumerate}

\section{DataSafe Design}
\label{sec:design}
\begin{figure}[!htbp]
\centering
\includegraphics[width=1.0\textwidth]{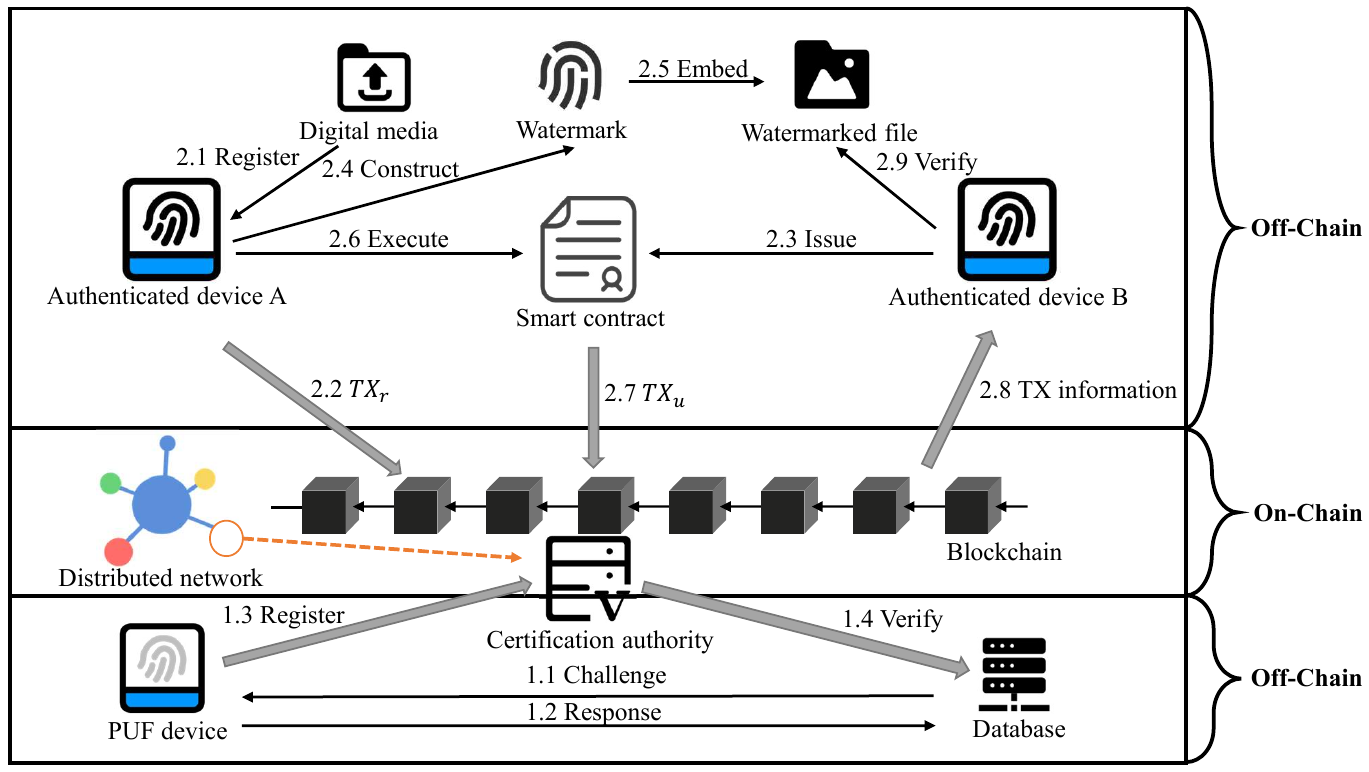}
\caption{DataSafe architecture} 
\label{fig:architecture}
\end{figure}
\subsection{Overview}
%In the previous section, we discussed the relevant background knowledge pertaining to copyright protection and traceability, as well as the security assumptions for using PUF devices. 
We tackle the challenges of identity binding and traceability during the transfer of copyright by proposing DataSafe, a copyright protection scheme for managing copyright transfer and traceability. The architecture of DataSafe is depicted in Figure \ref{fig:architecture}. In Section \ref{ss:regist}, we outline the application of security primitives for PUF devices to safeguard unique device keys and to record them on the blockchain. In Section \ref{ss:trans}, we elaborate on the mechanism of transferring copyright and authenticating digital watermarks between authorized devices. In Section \ref{ss:watermarking}, we elucidate the employment of improved spatial watermarking methods to integrate watermarks into digital media, and we address the potential secret leakage during the watermark extraction phase by utilizing shared keys. In Section \ref{ss:tracing}, we examine the measures taken to guarantee the accuracy of data traceability within the blockchain.

\subsection{PUF-based Device Registration}
\label{ss:regist}
Firstly, a PUF device is constructed within a secure setting to guarantee that the responses it produces in response to challenges are neither compromised nor counterfeited. Once the device is initialized, the manufacturer generates a random number $c$ and dispatches it to the device as a challenge. The device then creates a security primitive $\mathsf{DF}$ and auxiliary data $\mathsf{FE}$ on SRAM. The function $\mathsf{Gen}(\mathsf{DF},\mathsf{FE},c)$ obscures the random number $c$ to produce original output $o_{c}$. Subsequently, $\mathsf{Hash}(o_{c})$ is applied to derive a response $r$, which is then relayed back to the manufacturer. The manufacturer subsequently archives the device's identifier $\mathsf{id}$ assigned to the device during manufacturing and the challenge response pair $\langle c, r \rangle$ in a secure database, employing $\langle c, r\rangle$ as the device fingerprint for its authentication.

Next, the manufacturer selects a node within the blockchain to serve as the certification authority. This authority's role is limited to the registration of PUF devices. It connects to the secure database for device authentication and specifies the public parameters necessary for the generation of public-private key pairs. This includes the elliptic curve $E_{a, b}:y^3=x^3+ax+b$ over the finite field $F_{p}$, where $p$ is a large prime number, and $a,b$ meet the condition $4a^3+27b^2 \neq 0 \mod p, a, b\in F_{p}$. The authority then picks a large prime order $n$ and its corresponding generator $P\in E_{a,b}$ in $F_{p}$. The certification authority chooses a random number $\mathsf{sk}_{c}$ as the private key within the range of $n$, and computes the public key $\mathsf{pk}_{c}=\mathsf{sk}_{c}\cdot P$. The certification authority stores the private key and disseminates the public parameters and the public key.

Lastly, PUF devices need to register an identity on the blockchain. The device independently selects a random number $\mathsf{sk}\in n$ as its private key based on the predefined parameters, and computes $\mathsf{pk}=\mathsf{sk}\cdot P$ to create a public-private key pair $(\mathsf{pk},\mathsf{sk})$. Subsequently, the device generates origin output $o_{sk}$ based on $\mathsf{Gen}(\mathsf{DF},\mathsf{FE},\mathsf{sk})$, which is then saved in nonvolatile memory. The device then initiates an identity verification request to the certification authority. Through a key negotiation process, a temporary symmetric key $k$ is generated collaboratively by the certification authority and the device. The device provides its identifier $\mathsf{id}$ to the certification authority, which encrypts a random number $c$ using $k$ and forwards it to the device. Upon receiving $c$, the device decrypts it, generates $o_{c}$ based on $\mathsf{Gen}(\mathsf{DF},\mathsf{FE},\mathsf{c})$, and subsequently uses $\mathsf{Hash}(o_{c})$ to generate a response $r$. The device encrypts $r$ with $k$ and returns it to the certification authority. Upon receiving $r$, the certification authority checks the $\langle c, r\rangle$ pair against the entries in the secure database. If the verification check passes, the certification authority issues a acceptance notice to the device, registers it as an authenticated device on the blockchain, and archives the parameters $\langle \mathsf{id}, \mathsf{pk}\rangle$. The $\mathsf{id}$ is utilized by the certification authority for device identification, while the hash of $\mathsf{pk}$ serves as the device's transaction address $\mathsf{addr}$ on the blockchain. In the event that the verification does not pass, the certification authority dispatches a cancellation notice to the device, necessitating a re-initiation of the registration process.

\subsection{Copyright Registration and Transfer}
\label{ss:trans}
We leverage authenticated devices to act as secure vaults for copyrights. The device fingerprint of this device is embedded into the digital media. Copyright registration and transfer only occur after the device has been verified on the blockchain. This ensures that all participants in copyright transfer possess authenticated devices, guaranteeing identity cannot be forged and tamper-proof transaction records throughout the process.

\begin{figure}[!htbp]
\centering
\includegraphics[width=0.8\textwidth]{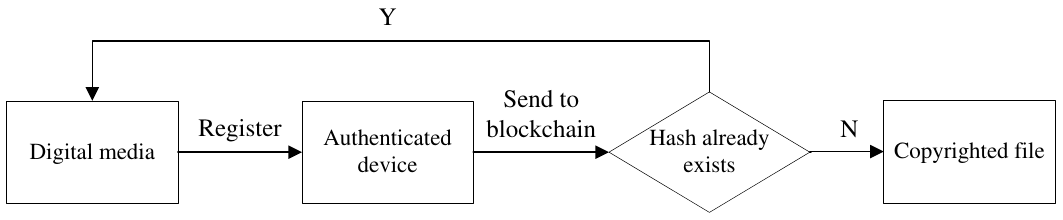}
\caption{Workflow of copyright registration} 
\label{fig:copyright_file_re}
\end{figure}

\subsubsection{Copyright Registration}
Before a digital media file can be acknowledged as copyrighted within the blockchain, it necessitates initial registration on the blockchain. The workflow of copyright registration is shown in the Figure \ref{fig:copyright_file_re}. This registration entails the authenticated device submitting a hash digest of the file onto the blockchain, which in turn creates a copyright marker. This copyright marker is composed of the file's hash digest $\mathsf{H}(f)$ and the blockchain address $\mathsf{addr}$ of the device holding the copyright.

Imagine a situation where digital media $f$ that is stored on an external drive needs to undergo copyright registration. The authenticated device $A$ computes the hash digest $\mathsf{H}(f)$ of $f$, encapsulates $\mathsf{H}(f)$ and $\mathsf{addr}$ into a transaction $\mathsf{TX}_{r}$ and dispatches it to the blockchain. Following this, after the transaction's validity is confirmed, miners within the blockchain network include the transaction in a block. Throughout this entire procedure, the original file remains securely with $A$, safeguarding it from any unauthorized access.

\begin{figure}[htbp]
\centering
\includegraphics[width=0.8\textwidth]{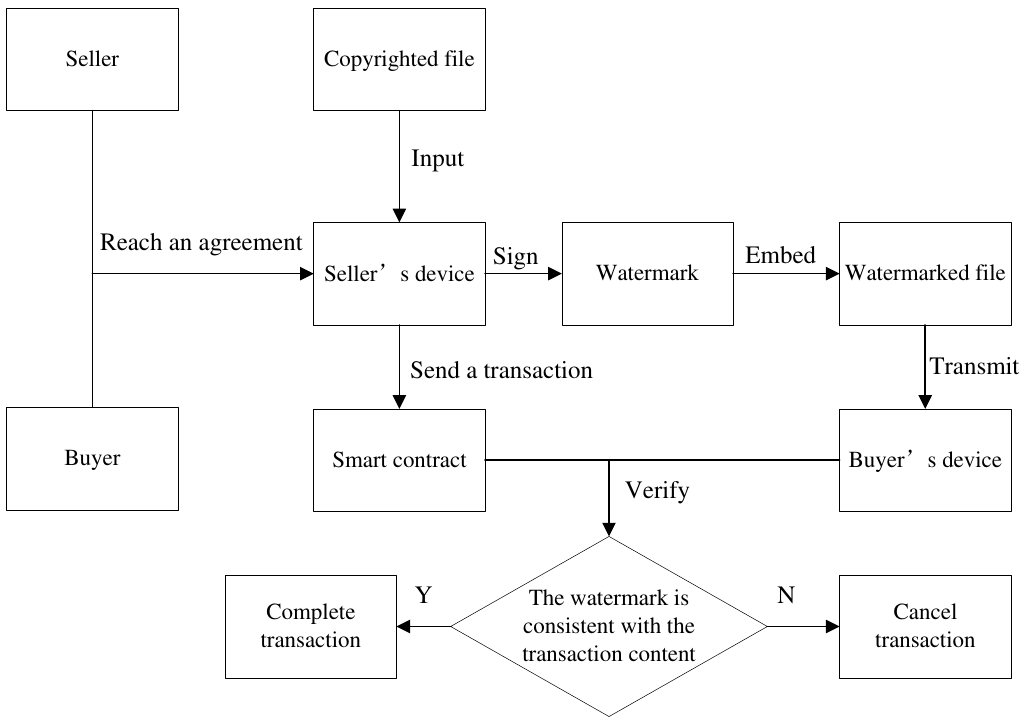}
\caption{Workflow of copyright transfer} 
\label{fig:copyright_tx}
\end{figure}

\subsubsection{Copyright Transfer}

As shown in Fig.~\ref{fig:copyright_tx}, transfer of copyrights in DataSafe consists of the following steps:

\indent 
\textit{Initiate transaction}: In this step, authenticated device $B$ intends to acquire $f$ from authenticated device $A$, leading to a negotiation between the two parties regarding the transaction specifics. This negotiation entails aspects such as the transaction price, deadlines for file delivery by $A$, payment deadline for $B$, and the deadline for confirming receipt. These negotiated terms are then documented within a smart contract, which is subsequently deployed onto the blockchain.

\textit{Pay for media}: $B$ transfers the agreed transaction price to the smart contract, which then notifies $A$ to initiate the delivery process.

\textit{Generate copyright transfer marker}: $A$ creates a copyright transfer marker $\mathsf{ctm}_{AB}$, which includes the transaction id of copyright marker and the addresses of both parties involved. Next, $A$ embeds a digital watermark into $f$ to produce a watermarked file $f'$ and generates the embedding position key $\mathsf{lk}$, as described in Section \ref{ss:watermarking}. Subsequently, $A$ packages the $\mathsf{H}(f')$ and the $\mathsf{ctm}_{AB}$ into a transaction $\mathsf{TX}_{u}$ and sends it to the smart contract, which then records the transaction on the blockchain.

\textit{Deliver media}: $A$ encrypts the embedding position key $\mathsf{lk}$ using $B$'s public key. Then $A$ sends $f'$ along with encrypted $\mathsf{lk}$ to $B$ via off-chain or peer-to-peer (P2P) communication.

\textit{Verification and transfer}: $B$ decrypts the received information to obtain $f'$ and $\mathsf{lk}$. First, $B$ calculates $\mathsf{H}(f')$ and compares it with the hash digest recorded on the blockchain. If the hashes do not match, $B$ cancels the transaction. If they do match, $B$ proceeds to extract the digital watermark from $f'$ using the method described in Section \ref{ss:watermarking}. Next, $B$ calculates the hash of $\mathsf{ctm}_{AB}$ and verifies the embedded watermark's correctness using $A$'s public key. After successful verification, $B$ sends a confirmation message to the smart contract, which then transfers the frozen tokens to $A$'s address.

\subsection{Embedding and Extraction of Watermarks}
\label{ss:watermarking}
\begin{figure}[!htbp]
\centering
\includegraphics[width=0.8\textwidth]{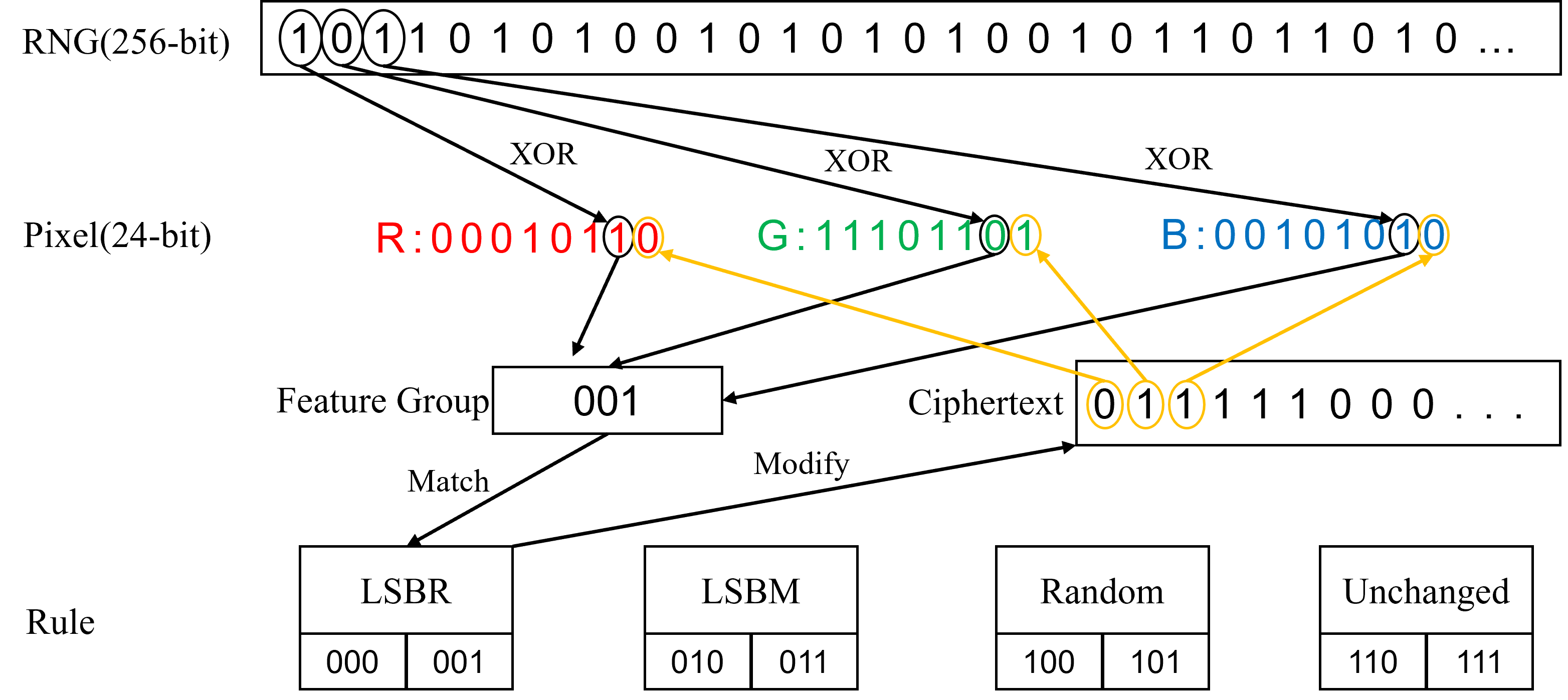}
\caption{Watermark embedding and location key matching} 
\label{fig:watermarking}
\end{figure}

The inherent security risk in digital watermarking arises from the requirement to share the key with the verifier. This presents a vulnerability because a verifier with malicious intent who has access to the key could potentially remove the watermark covertly. Furthermore, relying on a trusted third party for verification increases the risk of centralization. To counter these challenges, we propose a secure watermarking approach that moves the watermark embedding and extraction processes to the authenticated device itself.

To authenticate the identity of the copyright holder, we employ the Elliptic Curve Digital Signature Algorithm (ECDSA) to sign $\mathsf{ctm}_{AB}$ with a private key that is stored securely on the authenticated device. The signature produced is then transformed into a binary format to form a digital watermark $w_{AB}$, thereby ensuring the physical ownership of the digital media. Considering the performance limitations of the devices, we have focused on enhancing security. As a result, we opt for the Least Significant Bit (LSB) watermark embedding technique and generate a location key $\mathsf{lk}$ for embedding the watermark in the following manner.

As shown in Figure \ref{fig:watermarking}, the low bits stored in digital media are randomly distributed, with 0 and 1 each making up approximately half. This allows us to use the combination of the second lowest order bits as a condition for modifying the lowest order bits. For example, in a 24-bit true color image, the RGB format for a pixel $p_{ij}$ is represented as $R\{r_{0},r_{1},...,r_{7}\}$, $G\{g_{0},g_{1},...,g_{7}\}$, $B\{b_{0},b_{1},...,b_{7}\}$. We can extract the second lowest order bits $\{r_{6},g_{6},b_{6}\}$, use a random number generator to produce a random number $x$, and then perform a bitwise XOR operation with $\{r_{6},g_{6},b_{6}\}$ to obtain the feature group $\{x_{i}\oplus r_{6},x_{i+1}\oplus g_{6},x_{i+2}\oplus b_{6}\}$. We the classify these feature groups based on the size of watermark and predefined rules, defining the necessary modifications. Using an image as an example, we define four modifying actions for any given pixel.

\textit{Least Significant Bit Replacement(LSBR)}: Define a color channel $\mathsf{rgb}_{ij}$ of pixel $p_{ij}$, take a bit $m_{b}$ of ciphertext $m$, and replace the lowest bit of $\mathsf{rgb}_{ij}$ with $m_{b}$.
\begin{equation}
\mathsf{LSBR}(\mathsf{rgb}_{ij})=
\begin{cases}
    \mathsf{rgb}_{ij}, & if \ \mathsf{LSB}(\mathsf{rgb}_{ij})=m_{b}\\
    \mathsf{rgb}_{ij}+1, & if \ \mathsf{LSB}(\mathsf{rgb}_{ij})\neq m_{b} \ and \ \mathsf{rgb}_{ij}=0 \mod 2\\
    \mathsf{rgb}_{ij}-1, & if \ \mathsf{LSB}(\mathsf{rgb}_{ij})\neq m_{b} \ and \ \mathsf{rgb}_{ij}=1 \mod 2	
\end{cases}
\end{equation}

\textit{Least Significant Bit Matching(LSBM)}: the lowest bit of pixel $p_{ij-1}$ is filled with a random number. For pixel $p_{ij}$, define a color channel $\mathsf{rgb}_{ij}$, take a bit $m_{b}$ of ciphertext $m$, and match the lowest bit of $\mathsf{rgb}_{ij}$ with the $m_{b}$.
\begin{equation}
\mathsf{LSBM}(\mathsf{rgb}_{ij})=
\begin{cases}
    \mathsf{rgb}_{ij}, & if \ \mathsf{LSB}(\mathsf{rgb}_{ij})=m_{b}\\
    \mathsf{rgb}_{ij}+1, & if \ \mathsf{LSB}(\mathsf{rgb}_{ij})\neq m_{b} \ and \ \mathsf{rgb}_{ij}=0 \\
    \mathsf{rgb}_{ij}-1, & if \ \mathsf{LSB}(\mathsf{rgb}_{ij})\neq m_{b} \ and \ \mathsf{rgb}_{ij}=255 \\	
    \mathsf{rgb}_{ij}+rand, & if \ \mathsf{LSB}(\mathsf{rgb}_{ij})\neq m_{b} \ and \ 0<\mathsf{rgb}_{ij}<255 \\
\end{cases}
\end{equation}
Where $rand$ is a random number of 1 or -1.

\textit{Random number filling}: Fill the lowest bit of pixel $p_{ij}$ with random numbers.

\textit{Unchanged}: The lowest bit of pixel $p_{ij}$ remains unchanged.

The location key $\mathsf{lk}$ is constructed with $x$ and modifications. These modifications can be independently defined based on the file format, with the primary goal of enhancing security.

The verifier's authenticated device decrypts $\mathsf{lk}$ using a private key. It then processes the file according to the $x$ and the defined modifying actions within $\mathsf{lk}$ to extract $w_{AB}$. The device verifies the signature against the transaction on the blockchain to confirm copyright ownership. Since the malicious verifier lacks any knowledge about $\mathsf{lk}$, they cannot erase the watermark from the copyright file without damaging the file itself.

\subsection{Blockchain-based Copyright Tracing}
\label{ss:tracing}
The authenticated device $C$ proposes to acquire $f'$ from the authenticated device $B$, and they complete the transaction following the process described in Section \ref{ss:regist}. $B$ sends the watermarked file $f''$ and the location key $\mathsf{lk}'$ to $C$, recording the copyright transfer marker $\mathsf{ctm}_{BC}$ on the blockchain. $B$ encrypts $f'$ only when the hash digest matches $\mathsf{ctm}_{AB}$. Consequently, $C$ extracts the watermark using $t'$ and successfully verifies it, establishing the transfer direction of the copyright file as $A\Rightarrow B\Rightarrow C$. This ensures traceability of the copyright transfer.

\begin{figure}[!htbp]
\centering
\includegraphics[width=0.6\textwidth]{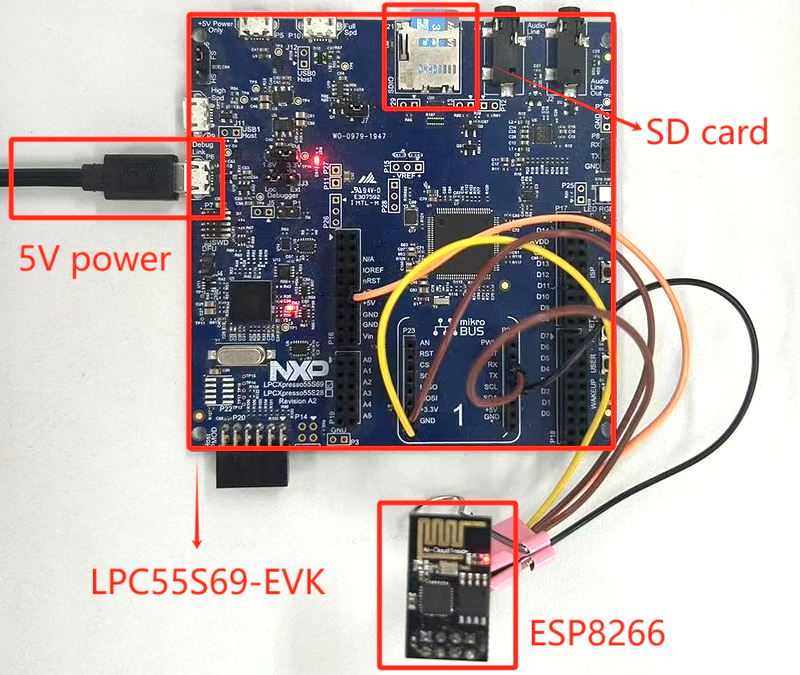}
\caption{Wiring diagram of DataSafe} 
\label{fig:wire diagram}
\end{figure}

\section{Prototype Implementation}
\label{sec:prototype}
In this section, we detail the development of a prototype for DataSafe, in line with our proposed scheme. The prototype was constructed using the LPC55S69-EVK development board, a product of NXP Semiconductor Company. This board is equipped with an LPC55S69 dual-core Arm Cortex-M33 microcontroller, which features a root of trust established through SRAM PUF and is compatible with Arm TrustZone technology. For further details on the LPC55S69-EVK, the user manual \cite{LPC55S6xMCU} is recommended for consultation.

We employed go-ethereum, a Golang-based implementation of the Ethereum protocol, as the foundational layer of our blockchain. A private blockchain was established across five computers, each furnished with an i7-7900@3.0GHz processor and endowed with 16 GB of memory, all operating on Ubuntu 18.04.1 GNU/Linux. Additionally, our configuration encompasses an authentication server.
We implemented DataSafe on the development board, which is energized by a 5V external power supply and linked to the blockchain through a WiFi module. The schematic wiring diagram of the system is illustrated in Figure \ref{fig:wire diagram}.

% \begin{figure}[!bht]
% {
% \begin{alltt}
% Transaction Hash:
%     0x5dc7b2f2acae6d3f129cbc2f1501f29a25ac5925c2415536eebf5e60d3b55755
% From: 0x8619bb67f62b09eed2aa597186680c6931d25e52
% To: null
% Block: 101
% TimeStamp: May-25-2024 09:05:35 GMT+0800 (CST)
% Data:
%     Prev-Transaction Hash: null
%     From:
%     04ab8793e998b9632590af11526d1b7b425783b3ff3d67bbe3fcdf1b65f335d15a
%       5e3841d334268f7faa82d5aace687c75af69ff054e11c920da1a8ed190d060a2
%     Hash:f9eab2fe9f516a7f07f881a896b32b50
% \end{alltt}
% }
% \caption{Register Transaction.} 
% \label{fig:register_tx}
% \end{figure}

\begin{figure}
\centering 
\begin{subfigure}{6cm}
{
%     \begin{alltt}
% Transaction Hash:
%     0x5dc7b2f2acae6d3f129cbc2f1501f29a25ac5925c24155
%     36eebf5e60d3b55755
% Data:
%     Prev-Transaction Hash: null
%     From: 0x8619bb67f62b09eed2aa597186680c6931d25e52
%     Hash: f9eab2fe9f516a7f07f881a896b32b50
%     \end{alltt}
    \includegraphics[width=\textwidth]{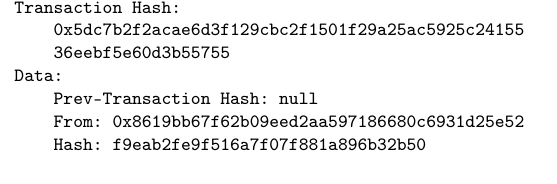}
}
\caption{Register Transaction}
\label{fig:register_tx}
\end{subfigure}
\begin{subfigure}{6cm}
{
%     \begin{alltt}
% Transaction Hash:
%     0x3725496d29fc94d6318750724f7f0424ad27e9102e4db4
%     4198a64421f8ab76bc
% Data:
%     Prev-Transaction Hash:
%     0x5dc7b2f2acae6d3f129cbc2f1501f29a25ac5925c24155
%     36eebf5e60d3b55755
%     From: 0x8619bb67f62b09eed2aa597186680c6931d25e52
%     To: 0x7b65b3b96dde01a0a4ae3e83d45c8a79d0ac8a69
%     New hash: a1f5b48e3cac1c823782f5f7b29a9fd7
%     \end{alltt}
    \includegraphics[width=\textwidth]{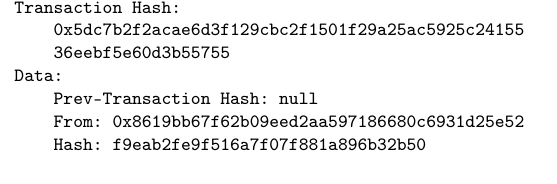}
}
\caption{Update Transaction}
\label{fig:update_tx}
\end{subfigure}

\caption{Transactions}
\label{fig:transactions}
\end{figure}

\begin{figure}[!bht]
\centering
    \begin{subfigure}{3cm}
        \includegraphics[width=\textwidth]{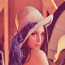}
        \caption{Origin image}
        \label{fig:orgin}
    \end{subfigure}
    \begin{subfigure}{3cm}
	\includegraphics[width=\textwidth]{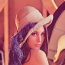}
	\caption{Covered image}
	\label{fig:covered}
    \end{subfigure}
    \caption{Watermarking results}
    \label{fig:lena}
\end{figure}

We kept the physical unclonable secrets of two boards within the authentication server. The boards confirmed their identities to the server by providing responses and were then duly registered on the blockchain. For our experimental purposes, we utilized a BMP-formatted image of Lena as a sample, registering it on the blockchain through the seller's board. The registration transaction is depicted in Figure \ref{fig:register_tx}.
Once an agreement was reached between the seller and the buyer, the buyer initiated a smart contract on the blockchain. On the board, the seller generated a copyright transfer marker and authenticated it with a private key. Subsequently, the watermark was integrated into the image, yielding the watermarked Lena image. Figures \ref{fig:orgin} and \ref{fig:covered} display the images before and after the watermark was embedded. We employed the Peak Signal to Noise Ratio (PSNR) to assess the imperceptibility of the embedded watermark, which measured 55.387 dB.
The seller's board proceeded to compute the hash digest of the image that now included the embedded watermark, crafted a transaction, and then transmitted it to the blockchain. This subsequent update transaction is portrayed in Figure \ref{fig:update_tx}. Upon acquisition of the file, the buyer decrypted the position key and authenticated the watermark with the aid of the board. Throughout this entire procedure, the position key remained undisclosed, ensuring that the buyer was unable to eliminate the watermark and gain access to the original image.

% \begin{figure}[!tbh]
% {\footnotesize
% \begin{alltt}
% Transaction Hash:
%     0x3725496d29fc94d6318750724f7f0424ad27e9102e4db44198a64421f8ab76bc
% From: 0x8619bb67f62b09eed2aa597186680c6931d25e52
% To:   0x7b65b3b96dde01a0a4ae3e83d45c8a79d0ac8a69
% Block: 111
% TimeStamp: May-25-2024 09:35:07 GMT+0800 (CST)
% Data:
%     Prev-Transaction Hash:
%     0x5dc7b2f2acae6d3f129cbc2f1501f29a25ac5925c2415536eebf5e60d3b55755
%     From:
%     04ab8793e998b9632590af11526d1b7b425783b3ff3d67bbe3fcdf1b65f335d15a
%       5e3841d334268f7faa82d5aace687c75af69ff054e11c920da1a8ed190d060a2
%     To:
%     042518ec20d46abdfafdbe8974227b4b02eeca184945b2032e40718bc538968e4a
%       8c3e96e4fa9ef7edf13a72905631661bbf808ffdd589a7b0f1d7b1425519c6f7
%     New hash:a1f5b48e3cac1c823782f5f7b29a9fd7
% \end{alltt}
% }
% \caption{Update Transaction} 
% \label{fig:update_tx}
% \end{figure}

In a situation where the seller partakes in deceitful conduct by selling files that have been registered on the blockchain on multiple occasions, the board has the capability to recognize these files by their identical hash digest and decline to register them. On the other hand, if the seller endeavors to register a file on the blockchain by tampering with the hash digest, the transfer process can indeed be finalized. Nevertheless, other users are able to pinpoint analogous content on the blockchain and ascertain the original physical proprietorship of the file by leveraging the timestamp and the capability to trace transactions. Subsequently, they can bring forth the authentication status of the seller's board to the blockchain, resulting in financial repercussions for the malevolent party.

\section{Conclusion}
\label{sec:conclusion}
The paper presents DataSafe, a copyright protection scheme that combines PUF and blockchain. It advocates for the secure storage of keys that represent physical devices within PUF devices and the integration of copyright transfer details into digital media through digital signatures. This approach solidifies the physical ownership aspect of digital copyrights. During the decryption phase, the embedded content is retrieved by merging the position key, and its authenticity is confirmed by referencing blockchain data, which guarantees the traceability of digital copyright. Both the embedding and extraction operations take place within the secure confines of the PUF device, safeguarding the digital watermark during the extraction process. In conclusion, the paper illustrates the practicality of the proposed scheme by showcasing a system prototype developed on the LPC55S69-EVK development board. Additionally, it addresses possible malicious distribution behaviors within the system and suggests appropriate countermeasures to mitigate such risks.

% \begin{credits}
% \subsubsection{\ackname} A bold run-in heading in small font size at the end of the paper is
% used for general acknowledgments, for example: This study was funded

% \subsubsection{\discintname}
% It is now necessary to declare any competing interests or to specifically
% state that the authors have no competing interests. Please place the
% statement with a bold run-in heading in small font size beneath the
% (optional) acknowledgments, for example: The authors have no competing interests to declare that are
% relevant to the content of this article. Or: Author A has received research
% grants from Company W. Author B has received a speaker honorarium from
% Company X and owns stock in Company Y. Author C is a member of committee Z.
% \end{credits}
\bibliographystyle{splncs04}
\bibliography{datasafe}

\end{document}